\documentclass[conference]{IEEEtran}

\usepackage[dvips]{graphicx}
\usepackage{float,makeidx,cite}
\usepackage{amssymb,amsmath,amsthm}

\providecommand{\abs}[1]{\lvert#1\rvert}
\newcommand{\ud}{\,\mathrm{d}}

\newtheorem{Theoi1}{Theorem}

\title{Throughput Efficient Large M2M Networks through Incremental Redundancy Combining}
\author{Amogh Rajanna and Mos Kaveh\\
ECE Department, University of Minnesota, USA. \{raja0088,mos\}@umn.edu}

\begin{document}
\maketitle
\begin{abstract}
In this paper, we investigate the performance of incremental redundancy combining as a new cooperative relaying protocol for large M2M networks with opportunistic relaying. The nodes in the large M2M network are modeled by a Poisson Point Process, experience Rayleigh fading and utilize slotted ALOHA as the MAC protocol. The progress rate density (PRD) of the M2M network is used to quantify the performance of proposed relaying protocol and compare it to conventional multihop relaying with no cooperation. It is shown that incremental redundancy combining in a large M2M network provides substantial throughput improvements over conventional relaying with no cooperation at all practical values of the network parameters.
\end{abstract}

 
\section{Introduction}
\label{new_intro}
M2M networks (Internet of Things) are envisioned as a key component of the future smart city solutions. A M2M network can provide services such as smart transportation, smart parking, environment/weather monitoring and smart grid. From a communication theoretic view, a M2M network for a smart city application consists of very large number of devices communicating with one another transporting information from source to destination over long distances. The performance of the relaying protocol used to forward information packets determines the reliability and efficiency of the network. In this paper, we propose a new relaying protocol for a large M2M network. The proposed protocol is expected to integrate with low power wide area network technologies such as Long Range (LoRa), SigFox, Ingenu etc.

In the literature, many opportunistic relaying protocols have been proposed with the key difference being the criterion used for selecting one relay from multiple relay candidates. A channel quality based criterion was proposed in \cite{Ganti} where the relay with the best channel to destination is selected. In \cite{Zhao}, the relay node which is closest to the destination is selected. In \cite{Baccelli}, the relaying protocol selects the relay farthest from the source in every hop. The protocol of \cite{Baccelli} had the salient features of a distributed relay selection procedure and the ability to transport packets between any two nodes in the network without prior connectivity. The work of \cite{Blomer_II} defines the progress rate density (PRD) of a network to study the relaying protocol of \cite{Baccelli}. The protocol is optimized to maximize the network throughput. 

Cooperative communication allows source and relays to utilize inherent space and time diversity leading to better throughput, outage and energy efficiency\cite{Andre}. In \cite{Zhao}, cooperative diversity in the form of incremental redundancy combining is applied to a source destination pair assisted by a fixed number of equidistant relays over a line. The codeword of a data packet is split into non overlapping blocks via puncturing and transmitted incrementally by the source and relays to the destination. The goal of our paper is to study the performance of incremental redundancy combining in a \textit{large M2M network to transmit information from source to destination over a long distance with large number of devices (nodes) acting as potential relays}. The network topology and wireless channel model of the present work has a big difference from that of \cite{Zhao}. The focus of \cite{Zhao} is on a single source destination pair over finite distance with equidistant relays on a line and the wireless channel has only fading, while we consider a large M2M network and the wireless channel has fading, path loss and interference, which is a crucial performance limiting factor.

The nodes of a large M2M network are modeled by a homogeneous Poisson Point Process (PPP). We propose to use incremental redundancy combining along with opportunistic relaying, and the resulting protocol is termed \emph{cooperative relaying protocol}. The performance of the cooperative relaying protocol is quantified through PRD of the network and compared to conventional relaying with no cooperation. Using an analytic approximation to the PRD, the protocol parameters are optimized. 
The gain in PRD due to cooperative relaying protocol is monotonic in the diversity order $M$, defined as \textit{the number of diverse transmissions that a destination node combines to decode a data packet}. For example, the gain in PRD at $\alpha=3$ from $M=1$ to $M=2$ is $26.5\%$ whereas from $M=2$ to $M=3$, the gain is $9.3\%$. As a function of path loss exponent $\alpha$, the cooperative relaying protocol has a consistent gain in PRD. For example, incremental redundancy combining with $M=2$ provides a gain of $26.5\%$ and $23.5\%$ at $\alpha=3$ and $\alpha=4$ respectively. 


\section{Cooperative Relaying for M2M Network} 
\label{M2M_relay}
Information packets are communicated from source to destination via a large number of isotropic hops through relays. Each source destination pair are apart by a large random distance and hence no predefined multihop path exists between them. For the model of large M2M network in our paper, the main focus is to transmit more information bits as far as possible in the source to destination direction per hop. The number of information bits is characterized by the transmission rate in the network.
The progress of an information packet is defined as the distance from the source in the source-destination direction over which the information bits are communicated. Spatial reuse is another key performance indicator. It is the ability to maintain simultaneous transmissions over different spatial regions of the network and is characterized by the density of transmissions in the network. Hence, the metric used in our paper for a large M2M network is \textit{Progress Rate Density} defined as \textit{the product of the number of information bits in bps communicated reliably per unit area of the network and the associated progress}. The PRD metric was stated in \cite{Blomer_II}. 

The process of transmitting data packets from source to destination by
incremental redundancy combining is explained in the following. Data packet is encoded into a codeword by the source node and split into non overlapping blocks through puncturing. Source transmits the $1^{st}$ block of the codeword which is received by potential relay nodes. Only a fraction of the potential relay nodes receiving the $1^{st}$ block of the codeword will decode the data packet based on the instantaneous SINR. It is assumed that every node in the network knows its own location and the data packet has information about source and destination locations. The relay nodes which decode the data packet compute the progress they offer. (Note that \emph{only} relays in between the source and destination have nonzero progress.)

The relays encode their progress into a $P$-bit vector $b_1b_2\cdots b_P$ and take part in a contention period of duration $P$ time units.
The activity in each of the $P$ time units is based exclusively on the $P$ bits of the progress bit vector. For every $0$ bit in the bit vector, the relay listens to the channel for the corresponding time unit and for every $1$ bit, the relay transmits a pulse. The contention activity of a relay proceeds in the following order of bits $b_P$ to $b_1$. For example, a relay having the bit vector $000110$ listens to the channel for first three time units, transmits two consecutive pulses and again listens to the channel. A relay quits the relay selection process only if it detects a pulse in the channel during a listening period, since it knows another relay has larger progress. At the end of contention period, the surviving relay has the most progress from the source and is selected as forwarding relay $1$ to transmit the $2^{nd}$ block of the codeword. If the source detects that no forwarding relay has been selected, then it retransmits the $1^{st}$ block and the selection procedure repeats. Since relays use LoRa (or SigFox), we assume that nodes can detect pulses emitted by other nodes over long distance during contention. 

When the forwarding relay $1$ transmits the $2^{nd}$ block of the codeword, all the potential relay nodes (and the destination) combine the $2^{nd}$ and $1^{st}$ block of the codeword to decode the data packet.
A fraction of the potential relay nodes have success based on the instantaneous SINR. The forwarding relay $2$ which transmits the $3^{rd}$ block of the codeword is selected based on the distributed contention scheme. Similarly when forwarding relay $M-1$ transmits block $M$ of the codeword, the relay nodes can combine $M$ blocks of the codeword for decoding. Every time a forwarding relay transmits a current block of the codeword, all potential relay nodes combine the current block with all $M-1$ previously received blocks of the codeword for decoding. The new forwarding relay selected transmits a block of the codeword which is complementary to all the $M-1$ recently received blocks. For both $i<M$ and $i\geq M$, the forwarding relay $i$ transmits the $q(i)+1$ block of the codeword, where $q(i)$=$\mod(i,M)$. For example if $M=3$, the first $M$ blocks of the codeword are transmitted by source and forwarding relays $1$ and $2$, respectively. Forwarding relays $4$ and $8$ transmit the $2^{nd}$ and $3^{rd}$ blocks of the codeword, respectively. 

This process of cooperative relaying where the relay nodes combine $M$ blocks of the codeword for decoding and distributively select the next forwarding relay which transmits a complementary block of the codeword continues until the data packet is successfully decoded at the destination. 

\section{System Model}
\label{sysmod}
We consider a wireless M2M network in which nodes are modeled by a 2-D homogeneous PPP $\Phi=\{i, X_i\}$, $i\geq 0$ of intensity $\lambda$, where $X_i$ denotes the coordinates of node $i$. The MAC layer uses the slotted ALOHA protocol. In every time slot, a node $i \in \Phi$ either acts as a transmit node with medium access probability (MAP) $p$ or as a receive node with probability $1-p$. The decision process to be either a transmit or receive node is independent from slot to slot and also of other nodes in the network. The parent PPP $\Phi$ can be split into 2 independent PPP's $\Phi^t$ and $\Phi^r$ of intensities $\lambda p$ and $\lambda(1-p)$ respectively. All nodes of $\Phi^t$ are either a source node or a forwarding relay node. All nodes of $\Phi^r$ are potential relay nodes. Each slot duration is split into two phases. In the $1^{st}$ phase, a node $\in \Phi^t$ transmits either its own or a data packet of another source node. In the $2^{nd}$ phase, all the nodes of $\Phi^r$ that decode the data packet by combining the current block of codeword from the $1^{st}$ phase (with, if any, the previous blocks of codeword received during previous slots) participate in the distributed contention scheme to select the forwarding relay for next hop. We assume i.i.d block fading across slots.

\begin{figure}[!hbtp]
\centering
\includegraphics[width=0.45\textwidth]{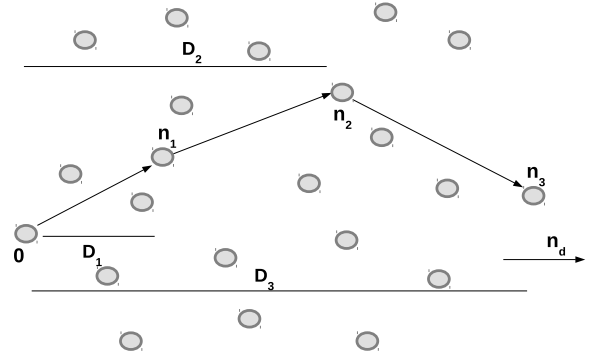}
\caption{M2M communication from source node $0$ to destination node $n_d$, which is at an asymptotic distance along the x-axis. Forwarding relay $i$ (node $n_i$) combines $i$ blocks of the codeword from nodes $\{0\cdots n_{i-1}\}$ for decoding and offers progress $D_i$ from origin along x-axis.}
\label{Ntw_snapshot_WCNC}
\end{figure}

Without loss of generality, we assume that node $0$ is the reference source. For simplicity, we consider the reference source to be located at the origin, i.e., $X_0=(0,0)$. Node $n_d$ is the reference destination, where $n_d$ is a large positive integer. It is located at an asymptotic distance along the x-axis, i.e., $X_{n_d}$ is a point on the positive x-axis at a large distance from the origin. The reference source destination pair is depicted in Fig.\ref{Ntw_snapshot_WCNC}. Conditioning on the source node at the origin does not affect the distribution of the homogeneous PPP $\Phi$ (See Slivnyak's theorem \cite{MartinBook} for more details). Source node at the origin encodes a data packet into a codeword, which is split into $M$ non-overlapping blocks  by puncturing. The source transmits the $1^{st}$ block of the codeword at code rate $R$. An important property of the puncturing process to note is that the $1^{st}$ block of the codeword is sufficient to decode the data packet.

The received signal at a node $v \in \mathbb{R}^2$ based on the transmission from the source node at origin is given by
\begin{equation}
\mathrm{y}=h_{0}|v|^{-\alpha/2}\mathrm{x}_0+\sum_{k\in\Phi^t}h_{k}|v-X_k|^{-\alpha/2}\mathrm{x}_k+\mathrm{z}\label{Rx_eq},
\end{equation}
where $h_{k}\sim \mathcal{CN}\left(0,1\right)$ is the Rayleigh fading coefficient from transmit node $k$, $\mathrm{x}_k$ is the message symbol of transmit node $k$ and $\alpha$ is the path loss exponent. In (\ref{Rx_eq}), the first term represents the desired signal, the second term represents the interference and $\mathrm{z}$ is the additive Gaussian noise. The SINR at receive node $v \in \mathbb{R}^2$ from the source node at origin is given by
\begin{equation}
\mathrm{SINR}\left(v,0\right)=\frac{\rho\abs{h_{0}}^2|v|^{-\alpha}}
{\sum_{k\in\Phi^t}\rho\abs{h_{k}}^2|v-X_k|^ {-\alpha}+\sigma^2},\label{SIR_eq}
\end{equation}
where $\sigma^2$ is the noise power and $\rho$ is the transmit power. In this paper, we focus on a M2M random network which has a large number of nodes. The network density will be in the interference limited regime where the effect of thermal noise is negligible. Hence in the following we assume $\sigma^2=0$ and the quantity in (\ref{SIR_eq}) becomes $\mathrm{SIR}\left(v,0\right)$.

All the nodes of $\Phi^r$ receive the $1^{st}$ block of the codeword from the source and the ones which decode the data packet participate in a distributed contention scheme. From the definition of progress in section \ref{M2M_relay}, the progress of a relay in the above presented system model is the distance from origin along the positive x-axis over which the information bits are communicated. Let the node $n_1\in \Phi^r$ with coordinate $X_{n_1}$ be the forwarding relay $1$. The relay selection is illustrated in Fig.\ref{Ntw_snapshot_WCNC}. The node $n_1$ offers the most progress from the origin among the relay nodes which decode the data packet using $1^{st}$ block of the codeword from node $0$. Mathematically, the progress offered by the node $n_1$ is given by
\begin{equation}
\mathrm{D}_1= \max_{i\in\Phi^r}\Big[\mathbf{1}\big(I_1\left(X_i\right)\geq R\big)~\abs{X_i}\cos\left(\theta\left(X_i\right)\right)\Big]
\label{hop1},
\end{equation}
where $I_1\left(X_i\right)=\log_2\left(1+\mathrm{SIR}\left(X_i,0\right)\right)$ is the mutual information (MI) achieved by relay node $i$ from node $0$, $\mathbf{1}\left(\cdot\right)$ is the indicator function and $\theta\left(\cdot\right)$ is the angle relative to the positive $x-$axis. As mentioned earlier, the destination node is at an asymptotic distance along the x-axis and hence the expression for $1^{st}$ hop progress in (\ref{hop1}) considers the progress offered by each relay node along the x-axis direction as measured by the $\abs{X_i}\cos\left(\theta\left(X_i\right)\right)$ term.

Since the node $n_1$ was able to decode the data packet, it will regenerate the $2^{nd}$ block of the codeword and transmit it in a future slot. In this paper, since the key focus is to measure how far the information bits are communicated from the source in the source-destination direction, we just assume that the forwarding relays transmit the blocks of the codeword within a few slots after they are selected.

During the $2^{nd}$ hop communication, the node $n_1$ transmits the $2^{nd}$ block of the codeword at rate $R$ in the $1^{st}$ phase of the slot it chooses to transmit. In the $2^{nd}$ phase of that slot, all the nodes $\in \Phi^r$  combine the $2^{nd}$ and $1^{st}$ blocks of the codeword from nodes $n_1$ and $0$, respectively. (Some nodes $\in \Phi^r$ may use only the $2^{nd}$ block without the $1^{st}$ block because they were not in receive mode when node $0$ was transmitting). Let node $n_2\in \Phi^r$ with coordinate $X_{n_2}$ be the selected forwarding relay 2. 
Similarly the node $n_M$ combines $M$ blocks of the codeword received from source and nodes $\{n_1, n_2, \cdots n_{M-1}\}$ for decoding and offers the most progress from origin. Mathematically, the progress from origin up to the node $n_M$ is given by
\begin{equation}
\mathrm{D}_M= \max_{i\in\Phi^r}\Bigg[\mathbf{1}\Big(\sum_{k=1}^M I_k\left(X_i\right)\geq R\Big)~\abs{X_i}\cos\left(\theta\left(X_i\right)\right)\Bigg] \label{prg_M},
\end{equation}
where $I_k\left(X_i\right)=\log_2\left(1+\mathrm{SIR}\left(X_i,X_{n_{k-1}}\right)\right)$ is the MI achieved by relay node $i$ from node $n_{k-1}$. Note that the node $n_i$ transmits the block $q(i)+1$ of the codeword. The cooperative relaying process continues until the data packet reaches the destination node $n_d$.

The performance of cooperative relaying protocol is compared to that of conventional relaying with no cooperation. Hence in the following, the performance metrics for relaying protocols with and without cooperation are defined. All forwarding relays transmit one block of the codeword at code rate $R$. The density of transmissions in the network is $\lambda p$ . The progress terms defined in (\ref{hop1}) and (\ref{prg_M}) are random variables and hence we define an expected measure of the same as
\begin{align}
d_M\left(R,p\right)&=\mathbb{E}\left[\mathrm{D}_M\right],
~~d_1\left(R,p\right)=\mathbb{E}
  \left[\mathrm{D}_1\right].
\label{aver_prg}
\end{align}
\subsubsection{No Cooperation (NC)}
For a conventional relaying protocol with no cooperation, the progress rate density of the network is given by
\begin{equation}
\mathrm{PRD}= R~\lambda p~d_1\left(R,p\right)\label{C1_exp}.
\end{equation}
\subsubsection{Incremental Redundancy Combining (IRC)}
$d_M$ in (\ref{aver_prg}) is a measure of progress which spans $M$ hops. To compare the PRD of cooperative relaying protocol to that in (\ref{C1_exp}), we need a measure of progress per hop. So we define $d_M-d_{M-1}$ as the progress per hop when the relay nodes combine $M$ blocks of the codeword. Hence for the cooperative relaying protocol, progress rate density of the network is given by
\begin{equation}
\mathrm{PRD}= R~\lambda p~\big(d_M\left(R,p\right)- d_{M-1}\left(R,p\right)\big)\label{C2_exp}.
\end{equation}
In the next section, we optimize the cooperative relaying protocol by maximizing the PRD metric in (\ref{C2_exp}).

\section{Protocol Optimization}
\label{opti_prd}

The PRD in (\ref{C1_exp}) and (\ref{C2_exp}) are evaluated based on  simulation. In this section, we seek to optimize the cooperative relaying protocol by developing an analytic approximation to PRD of the network and optimizing the analytic function. It is conceptually infeasible to evaluate the distribution and expectation of the progress $D_M$ defined in (\ref{prg_M}). Alternatively we develop a heuristic approximation to the expected progress $d_M$. The approximation is based on the concept of decoding cells introduced in \cite{Bacc_book1}. Decoding cells in their simplest form are areas in $\mathbb{R}^2$ containing points with successful decoding of packets transmitted from the origin.

A decoding cell for incremental redundancy combining $\mathrm{\Sigma}_M$ is defined as
\begin{IEEEeqnarray}{rCl}
\mathrm{\Sigma}_M&=&\left\lbrace v \in \mathbb{R}^2:~\sum_{k=1}^M I_k \geq R \right\rbrace\label{cell_defM}\\
I_k&=&\log_2\left(1+\mathrm{SIR}\left(v,\eta_{k-1}\right)\right),~
\eta_{k-1}=\left(\tilde{d}_{k-1},0\right)\label{Ek1}.
\end{IEEEeqnarray}
To better understand $\mathrm{\Sigma}_M$, let us consider $M=2$ with  the point $\eta_0$ being origin. For the point $\eta_1$, $\tilde{d}_1$ is an approximation to the expected progress $d_1$ in (\ref{aver_prg}). $\tilde{d}_1$ has a closed form expression as a function of system parameters but for the ease of presentation, the expression is presented later.

The cell $\mathrm{\Sigma}_2$ contains all $v \in \mathbb{R}^2$ that decode the packet using two blocks of the codeword from origin and $\eta_1$, respectively. The point $\eta_1$ in (\ref{Ek1}) represents the equivalent of the location of forwarding relay $1$. Although the progress in (\ref{prg_M}) involves the instantaneous random location of forwarding relay $1$, we use an approximate expected location given by $\eta_1$ in the definition of cell $\mathrm{\Sigma}_2$ for analytical tractability. The coordinate-1 of forwarding relay $1$ is given in (\ref{hop1}). Since we are interested in the expected location of the forwarding relay $1$ for cell definition, we set the coordinate-1 of $\eta_1$ to $\tilde{d}_1$. There is no information about the coordinate-2 of forwarding relay $1$ in (\ref{hop1}). However we are only interested in the progress from origin along the positive $x$-axis. Hence using PPP stationarity to simplify the analysis, we set the coordinate-2 of $\eta_1$ to $0$. Such a point $\eta_1$ will be useful to compute a tractable and valuable approximation to the expected progress $d_2$.

The average cell area is given by
\begin{equation}
\mathbb{E}\big[\abs{\mathrm{\Sigma}_{M}}\big]=\int_{\mathbb{R}^2}
\mathbb{P} \Big(\sum_{k=1}^M I_k \geq R\Big)~\ud v \label{are_eq}.
\end{equation}
An interpretation of the average cell area is that it contains all $v \in \mathbb{R}^2$ which in the expected sense can decode the data packet using $M$ blocks of the codeword. By homogeneity of the PPP $\Phi$, the relay nodes in the average cell area are uniformly distributed. Using these properties, the following theorem derives an analytic approximation to the expected progress for incremental redundancy combining $d_M\left(R,p\right)$ in (\ref{aver_prg}).

\begin{Theoi1}
\label{The_1}
An approximation to the expected progress of cooperative relaying protocol with incremental redundancy combining $\tilde{d}_M\left(R,p\right)$ is given by
\end{Theoi1}
\begin{IEEEeqnarray}{rCl}
\tilde{d}_M\left(R,p\right)&=& \frac{\sqrt{\abs{W_M}}+\tilde{d}_{M-1}}{2}~ \left(1-\frac{1-e^{-c_M}}{c_M}\right)\label{dM_appr}\\
c_M&=&\frac{\lambda (1-p)}{2}\left(\abs{W_M}+\tilde{d}_{M-1} \sqrt{\abs{W_M}}\right),
\end{IEEEeqnarray}
where $\abs{W_M}$ in (\ref{dM_appr}) is equal to the average cell area in (\ref{are_eq}).
Since $\tilde{d}_M$ is recursive, an expression for $\tilde{d}_1$ is essential to complete Theorem 1. The $\tilde{d}_1$ expression is derived based on a decoding cell with only origin as the center. An expression for $\tilde{d}_1$ follows from (\ref{dM_appr}) with $M=1$ and using $\tilde{d}_0=0$. From (\ref{are_eq}), $\abs{W_{1}}$ is given by
\begin{align}
\abs{W_{1}}&\stackrel{(a)}{=}\frac{\pi}{\lambda
p G\left(\alpha\right)\left(2^R-1\right)^{\delta}},
\end{align}
where $G\left(\alpha\right)=\tfrac{\pi\delta}{\sin\left(\pi\delta\right)}$, $\delta=2/\alpha$ and (a) follows by using in (\ref{are_eq}), $\mathbb{P}\left(I_1 \geq R\right)=\exp(-\pi \lambda p G\left(\alpha\right)\left(2^R-1\right)^{\delta}\abs{v}^2)$ from \cite{Baccelli}.
\begin{IEEEproof}
Refer to Appendix \ref{sec:ProofTheo1} for derivation of (\ref{dM_appr}) and computation of $\abs{W_M}$.
\end{IEEEproof}
%

For optimal operation of the cooperative relaying protocol,
coding rate $R$ and MAP $p$ need to be optimized.
Maximization of the PRD in (\ref{C2_exp}) is given by
\begin{equation}
\langle R,p\rangle=\arg \max_{R,p}~~R~\lambda p~\big(d_M\left(R,p\right)- d_{M-1}\left(R,p\right)\big)\label{Opt_va}.
\end{equation}
Both optimal $R$ and $p$ are solved by simulation and the numerical results are presented in section \ref{sim_res}. The heuristic  approximation $\tilde{d}_M$ in (\ref{dM_appr}) is very valuable in that it helps to solve the optimization in (\ref{Opt_va}) analytically. 
The analytic approximation to optimal $R$ and $p$ is given by
\begin{equation}
    \langle \tilde{R},\tilde{p}\rangle=\arg \max_{R,p}~~ R~\lambda p~\Big[\tilde{d}_M\left(R,p\right)- \tilde{d}_{M-1}\left(R,p\right)\Big]\label{Apr_va}.
\end{equation}
The objective function is concave and the KKT points are solved by gradient descent methods. Note that for no cooperation case ($M=1$), both $d_{M-1}$ and $\tilde{d}_{M-1}$ are set to $0$.

\section{Numerical Results}
\label{sim_res}
We present numerical results illustrating the performance of the cooperative relaying protocol proposed in the paper, which in practice sits on top of LoRa, SigFox, Ingenu etc. The performance is measured by simulating the reference source destination communication. The values of network parameters used in the simulation are $\lambda=1$ and $\alpha\in[2.5,4]$. In the numerical results, we also include the performance of another cooperative relaying scheme known as repetition combining (RC). In RC, the SIRs of the different transmissions add up at the relay (destination). This is similar to Chase combining. The analytical results of the paper are also applicable to this scheme with the key difference being that in (\ref{prg_M}), (\ref{cell_defM}) and (\ref{are_eq}), the SIRs add up rather than the MI terms.
\begin{figure}[!hbtp]
\centering
\includegraphics[width=0.45\textwidth]{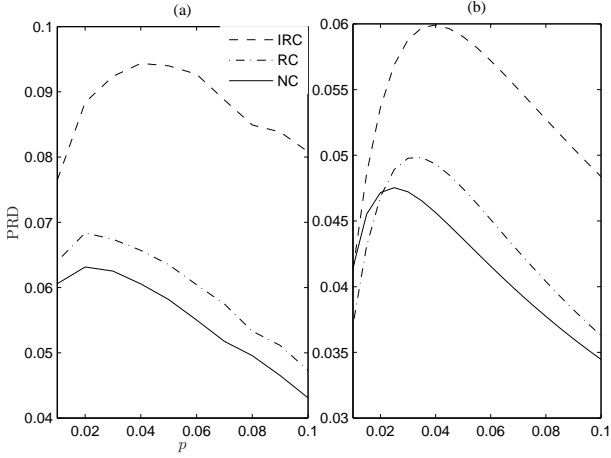}
\caption{PRD plotted against MAP $p$ for relaying protocols with NC, RC and IRC, respectively at $R=3$ and $\alpha=3$. The curves are plotted based on (\ref{C1_exp}) and (\ref{C2_exp}). For RC and IRC, the diversity order is $M=2$. (a) Simulation and (b) Analytic.}
\label{Coop_Relay_benefit}
\end{figure}

Fig. \ref{Coop_Relay_benefit} shows a plot of the PRD as a function of the MAP $p$ at $R=3$ and $\alpha=3$. In conventional relaying with NC, each relay node can use \emph{only} the transmission from the current forwarding relay for decoding. But for both IRC and RC with $M=2$, the relay nodes will have access to the transmissions from both the current and the previous forwarding relays. The relay nodes combine the two transmissions for decoding, thus extracting the space and time diversity inherent in the network leading to a higher throughput compared to the relaying protocol with NC. In IRC, every forwarding relay supplies new parity symbols to decode the data packet. These new parity symbols in addition to the available space time diversity enable the relay nodes to decode more information bits and thus achieve a higher PRD compared to RC. This effect is observed in the plotted curves. 

\begin{figure}[!hbtp]
\centering
\includegraphics[width=0.45\textwidth]{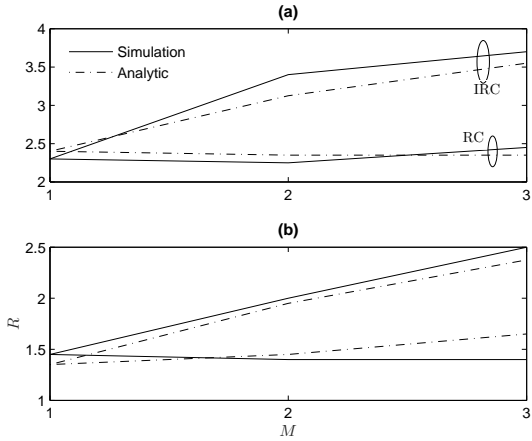}
\caption{The optimal $R$ from (\ref{Opt_va}) and (\ref{Apr_va}) plotted against the diversity order $M$ at a) $\alpha=4$ and b) $\alpha=3$.}
\label{Coop_Relay_Rop_M}
\end{figure}
In Fig. \ref{Coop_Relay_benefit}, both the simulation and theoretic curves have the same optimal MAP $p$. The analytical results in Theorem \ref{The_1} facilitate to obtain the optimal $R$ and $p$ for efficient network operation without extensive simulation. This point is further explored in Fig. \ref{Coop_Relay_Rop_M}, which plots the optimal $R$ as a function of $M$. The maximization of PRD is based on (\ref{Opt_va}) and (\ref{Apr_va}). First, the optimal $R$ and $p$ for NC ($M=1$) are obtained. For IRC, increasing $M>1$ leads to more MI accumulation and hence the rate $R$ is optimized while keeping the $p$ fixed to the optimal $p$ for NC. The optimal $R$ increases monotonically with $M$ for both $\alpha=3$ and $\alpha=4$. However for RC, both $R$ and $p$ are optimized.

\begin{figure}[!hbtp]
\centering
\includegraphics[width=0.45\textwidth]{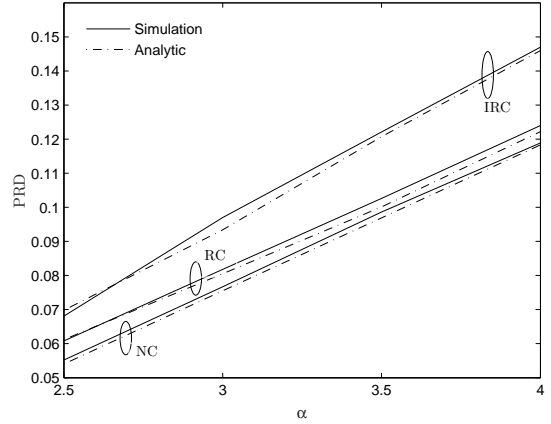}
\caption{Maximal PRD against the path loss exponent $\alpha$ for relaying protocols with NC, RC and IRC as per (\ref{Opt_va}) and (\ref{Apr_va}). For RC and IRC, $M=2$.}
\label{Coop_Relay_PRD_alpha}
\end{figure}
Fig. \ref{Coop_Relay_PRD_alpha} shows a plot of the maximized PRD values against the path loss exponent $\alpha$ for relaying protocols with NC, RC and IRC. Cooperative relaying in the form of IRC leads to a near constant gain in network throughput at varying values of $\alpha$. From the curve for IRC with $M=2$, the network has a $26.5\%$ gain in PRD at $\alpha=3$. At $\alpha=4$, the gain in PRD is $23.5\%$. As $\alpha$ decreases, the effect of interference in the network increases and SIR decreases, enhancing the benefit of doing cooperative relaying. This nature of variation of the PRD gain as a function of $\alpha$ is also valid for RC. 

\begin{figure}[!hbtp]
\centering
\includegraphics[width=0.45\textwidth]{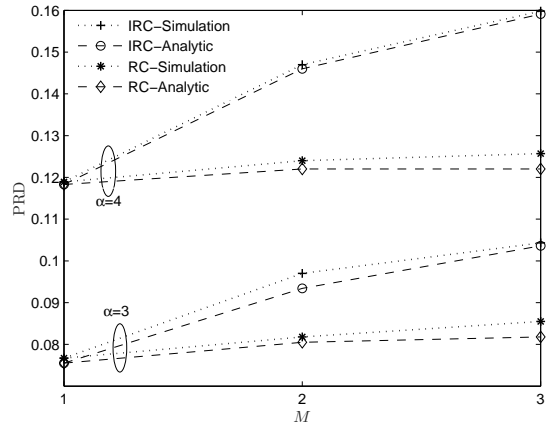}
\caption{Maximal PRD of the network from (\ref{Opt_va}) and (\ref{Apr_va}) plotted against diversity order $M$ at $\alpha=\{3,4\}$.}
\label{Coop_Relay_PRD_M}
\end{figure}
Fig. \ref{Coop_Relay_PRD_M} shows a plot of the PRD as a function of the diversity order $M$ for both RC and IRC at $\alpha=\{3,4\}$. The PRD values are maximized based on (\ref{Opt_va}) and (\ref{Apr_va}). From the curve for IRC in Fig. \ref{Coop_Relay_PRD_M}, it is observed that at $\alpha=3$ the PRD increases by $26.5\%$ when the diversity order changes from $M=1$ to $M=2$ but when $M$ goes from $M=2$ to $M=3$, the PRD gain is only $9.3\%$. For cooperative relaying with $M>2$, the signal strengths, i.e., SIR of the different transmissions that a relay node combines to decode a packet are non-identical. For example, consider the reference source destination communication when $M=3$. The relay node $n_3$ combines three transmissions from the forwarding relay nodes $\{n_2, n_1, 0\}$ which are of decreasing signal strength on average due to the increasing distance from $n_3$. As a result of the decreasing signal strength of the diverse transmissions, the benefit of cooperative relaying in terms of PRD gain becomes monotonic with the diversity order $M$. Such a monotonic nature of increase of the PRD with $M$ is consistent at $\alpha=4$ and also holds true for RC. 

The PRD values of the network from simulation when operated at the $R$ and $p$ given by (\ref{Opt_va}) and (\ref{Apr_va}) are shown in Figs. \ref{Coop_Relay_PRD_alpha} and \ref{Coop_Relay_PRD_M}. Based on the proximity of the PRD values derived from simulation based optimization and analytic optimization in both figures, we observe that the analytic optimization in (\ref{Apr_va}) operates the network very close to the optimal PRD point obtained from (\ref{Opt_va}).
\section{Conclusion}
\label{con_cl}
In this paper, a low complexity cooperative relaying protocol based on incremental redundancy combining for a large M2M network was presented. An analytic approximation to the PRD of the network was used to optimize the performance of cooperative relaying protocol. The optimized protocol leads to substantial gain in PRD over conventional relaying with no cooperation at key scenarios of the network parameters. For the large M2M network model, the PRD gain is consistent at all values of path loss exponent $\alpha$ and monotonic increasing in diversity order $M$. The presented numerical results emphasize the potential to provide a more efficient and reliable PHY-MAC and enhance the performance of SigFox, LoRa etc for a large M2M network (urban IoT).

\appendices

\section{Proof of Theorem 1}
\label{sec:ProofTheo1}
\begin{figure}[!hbtp]
\centering
\includegraphics[width=0.45\textwidth]{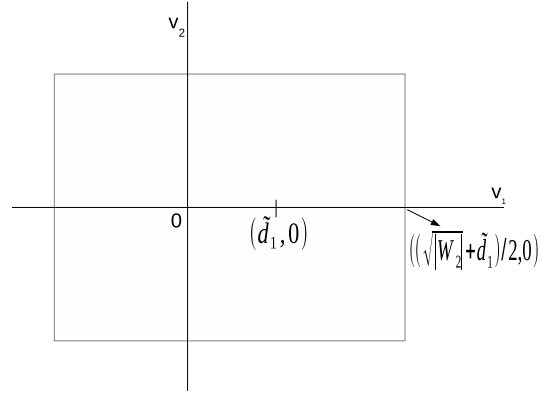}
\caption{A square $W_2$ centered around the two points origin and $\eta_1=(\tilde{d}_1,0)$ represents the decoding cell area. The center of the square is $(\tilde{d}_1/2,0)$. Maximum progress offered by nodes of $\Phi^r$ in $W_2^+$ is $(\sqrt{\abs{W_2}}+\tilde{d}_1)/2$.}
\label{Cell_Diag}
\end{figure}

We first derive the proof by considering $M=2$ and a simple extension yields $M>2$. Since a square has a simple relation between its area and  length, consider a square $W_2$ with area $\abs{W_2}=\mathbb{E}\big[\abs{\mathrm{\Sigma}_2}\big]$ centered around the two points, origin and $\eta_1$ as shown in Fig. \ref{Cell_Diag}. Let $W_2^+$ represent the portion of $W_2$ in the positive $v_1$ axis. The area of $W_2^+$ is given by $\abs{W_2^+}=\sqrt{\abs{W_2}}\cdot J$, where $J=(\sqrt{\abs{W_2}}+\tilde{d}_1)\big/2$.

Define $K$ as the number of nodes of $\Phi^r$ in $W_2^+$. For stationary PPP $\Phi^r$, $K=\abs{\Phi^r\left(W_2^+\right)}$ is Poisson distributed with parameter $c_2=\lambda (1-p)\abs{W_2^+}$. The nodes of $\Phi^r$ in $W_2^+$ offer a maximum progress of $J$. Hence based on the above mentioned properties, an approximate expression for $\mathbb{E}\big[\mathrm{D}_2\big]$ is given by
\begin{align}
\tilde{d}_2\left(R,p\right)&= \sum_{k=0}^{\infty}\mathbb{E}\Big[\max_{i\leq k}~U_{i,v_1}\big | K=k\Big]~ 
\mathbb{P}\left(K=k\right)\label{int_eq_prg}
\end{align}
\begin{equation}
=\sum_{k=0}^{\infty} \frac{\sqrt{\abs{W_2}}+\tilde{d}_1}{2}~ \frac{k}{k+1}~\mathbb{P}\left(K=k\right)\nonumber	
\end{equation}
\begin{equation}
=\frac{\sqrt{\abs{W_2}}+\tilde{d}_1}{2}~\sum_{k=0}^{\infty}
\mathbb{P}\left(K=k\right)\left(1-\frac{1}{k+1}\right)\nonumber	
\end{equation}
\begin{equation}
=\frac{\sqrt{\abs{W_2}}+\tilde{d}_1}{2}~\left(1-\frac{1-e^{-c_2}}{c_2}\right)\label{cell_appr},	
\end{equation}
where in (\ref{int_eq_prg}), $U_{i,v_1}$ is the coordinate-1 of 
relay $i$ uniformly distributed over $\left[0,J\right]$. Due to space constraints, we omit the full steps to arrive at (\ref{cell_appr}). (See \cite{Rajanna} for details). 

For $M>2$, the proof is based on the same lines with square $W_M$ centred around origin and $\eta_{M-1}$. To complete the proof, we mention that the expression for the square area $\abs{W_M}$, i.e., $\mathbb{E}\big[\abs{\mathrm{\Sigma}_M}\big]$ in (\ref{are_eq}) is computed by numerical integration. However for $M=2$, an efficient lower bound for $\abs{W_2}$ is given in \cite{Rajanna}.
\bibliography{References_II}
\bibliographystyle{IEEEtran}
\end{document}